\documentclass[preprint,showkeys,secnumarabic,amsmath,amssymb]{revtex4}
\usepackage{graphicx}

\begin{document}

  \title{Probing Lambda-DGP Braneworld Model}
   \author{A. Ravanpak}
\email{a.ravanpak@vru.ac.ir}
\affiliation{Department of Physics, Vali-e-Asr University, Rafsanjan, Iran}
\author{H. Farajollahi}
\email{hosseinf@guilan.ac.ir}
\affiliation{Department of Physics, University of Guilan, Rasht, Iran}
\affiliation{School of Physics, University of New South Wales, Sydney, NSW, 2052, Australia}
\author{G. F. Fadakar}
\email{gfadakar@guilan.ac.ir}
\affiliation{Department of Physics, University of Guilan, Rasht, Iran}

\date{\small {\today}}

\begin{abstract}
In this article we study cosmic dynamics in the context of normal branch of DGP braneworld model. Using the current Planck data, we best fit the model and cosmological parameters in non-flat $\Lambda$DGP. With the transition redshift as a basic variable and statefinder parameters, our result shows that the Universe starts its accelerated expansion phase, slightly earlier than expected in $\Lambda$CDM cosmology. The result also alleviates the coincidence problem of the $\Lambda$CDM model.
\end{abstract}

   \maketitle

\section{Introduction}
\label{sect:intro}

Nowadays, we know from observations that our Universe is experiencing an accelerated expansion phase, with some unknown mechanism \cite{Riess98},\cite{Per98},\cite{Per99},\cite{Kowalski08},\cite{Amanullah10},\cite{Sahni00}. A mysterious fluid with negative pressure dubbed dark energy is the most important candidate where can drive this acceleration \cite{Peebles03},\cite{Padmanabhan03},\cite{Oliveros15},\cite{Brevik15},\cite{Frieman08},\cite{Wang07}. The cosmological constant, $\Lambda$, is the simplest dark energy component despite of the related coincidence problem \cite{Weinberg89},\cite{Zaripov14},\cite{Basilakos1},\cite{Basilakos10}. Also, we know that the Universe underwent a phase transition during its evolution. At this era it has changed its decelerated expansion to an accelerated phase. Thus, the so called deceleration parameter, $q(z)$, switches from positive to negative values for a specific value of redshift called transition redshift, $z_t$.

Besides, we expect the redshift of any cosmological source change after a time interval because of the evolution and expansion of the Universe. Although this redshift drift is small and could not be measure at low redshifts, it is the unique way to determine the expansion history of the Universe, directly and without any defaults \cite{Sandage62},\cite{McVittie62}. Observationally, we can express this change in redshift as a spectroscopic velocity drift of the source which is in order of several cm s$^{-1}$ yr$^{-1}$.

On the other hand, in the past two decades the theory of extra dimensions has attracted a great deal of attention between researchers \cite{Maia04},\cite{Rudra11},\cite{Shahidi11}. It thrived by braneworld scenario of Randall and Sundrum (RS) and developed gradually. Among many extensions of RS model, the model proposed by Dvali, Gabadadze and Porrati (DGP) is of particular interest, because one branch of this model explains the certain late-time acceleration of the Universe without any dark energy component, irrespective of its own problems.
In DGP braneworld model, our 4D world is a brane embedded in an infinite 5D Minkowskian bulk. Also, all the matter fields are confined to the brane and only gravity can leakage into the bulk. According to the two ways that brane can be embedded in the bulk, the model features two separate branches denoted by $\epsilon=\pm1$, with distinct characteristics. The $\epsilon = +1$ branch is called the self-accelerating solution in which the Universe experiences a late-time acceleration due to modification of gravity. On the other hand the $\epsilon = -1$ branch where is not able to accelerate without a dark energy component is called the normal branch \cite{Rham08},\cite{Dvali00},\cite{Lopez09},\cite{Quiros09}.

This paper is organized as follows. In section II, we study the $\Lambda$DGP model, i.e., a spatially non-flat DGP model in the presence of a cosmological constant as a dark energy component. We express the transition redshift in terms of our model parameters via $q(z_t)=0$ and then estimate it numerically, using a best-fitting procedure. Section III, deals with the redshift drift and drift velocity. In this section we test our model with observations. Also, in each case we compare our results with the $\Lambda$CDM model. In section IV, a statefinder diagnostic procedure is used to distinguish our model among different dark energy models. In section V, the important coincidence problem is investigated in our model. Section VI includes conclusion and remarks.

\section{TRANSITION REDSHIFT}

As we mentioned in introduction our Universe has experienced a transition from a decelerated expansion to an accelerated one, in its expanding history. The redshift of this transition is called transition redshift, $z_t$, which is one of the important parameters in cosmology. In this section we are trying to obtain this value directly, using a numeric approach in non-flat $\Lambda$DGP model. To this aim, we start by Friedmann equation and after calculating the related deceleration parameter, we use the condition $q(z_t)=0$, to obtain an expression for $z_t$, in terms of our model parameters. Then, considering $z_t$, as a free parameter, a best-fitting procedure is used to determine the best values of model parameters. Using these values we plot the curves $q(z)$ and ($z_t,\Omega_{m0}$), for the model under consideration and compare them with $\Lambda$CDM model. Also, we can compare the curve ($z_t,\Omega_{m0}$), with the observational constraints from Planck satellite.

In non-flat $\Lambda$DGP model we have the following modified Friedmann equation on the brane \cite{Xu10}
\begin{eqnarray}\label{fried33}
H^2+\frac{K}{a^2}&=&(\sqrt{\frac{\rho}{3M_p^2}+\frac{1}{4r_c^2}}-\frac{1}{2r_c})^2,
\end{eqnarray}
where $H=\dot a/a$, $a=a(t)$, $M_p$ and $r_c=\kappa^2_{(5)}/\kappa^2_{(4)}$, are the Hubble parameter, scale factor, the 4D Planck mass and the so called crossover distance, respectively. Also, $\rho=\rho_m+\rho_\Lambda$, where $\rho_m$ is related to the matter content on the brane and $\rho_\Lambda=M_p^2\Lambda$. Also, $K=\pm1$, is the curvature parameter.
Using the fractional energy densities as
\begin{equation}\label{dimlesspar}
\Omega_{m}=\frac{\rho_{m}}{3M_p^2H^2}, \quad \Omega_{\Lambda}=\frac{\Lambda}{3H^2}, \quad \Omega_{r_c}=\frac{1}{4r_c^2H^2}, \quad \Omega_{K}=-\frac{K}{a^2H^2},
\end{equation}
one can rewrite the Friedmann equation as below
\begin{equation}\label{fried4}
E^2(z)=(\sqrt{\Omega_{m0}(1+z)^3+\Omega_{r_c0}+\Omega_{\Lambda0}}-\sqrt{\Omega_{r_c0}})^2+\Omega_{K0}(1+z)^2.
\end{equation}
Here, $E(z)=H(z)/H_0$ and the zero index means the present value of any cosmological parameters.
The deceleration parameter, $q=-\ddot a/(aH^2)$, in terms of redshift is defined as
\begin{equation}\label{q}
q(z) = \frac{(1+z)}{H(z)}\frac{dH}{dz}-1.
\end{equation}
So, it can be rewritten in terms of fractional energy densities as
\begin{eqnarray}
  q(z) &=& \frac{3\Omega_{m0}(1+z)^3(\sqrt{\Omega_{m0}(1+z)^3+\Omega_{\Lambda0}+\Omega_{r_c0}}-\sqrt{\Omega_{r_c0}})}
  {2({(\sqrt{\Omega_{m0}(1+z)^3+\Omega_{\Lambda0}+\Omega_{r_c0}}-\sqrt\Omega_{r_c0})^2+\Omega_{K0}(1+z)^2)}\sqrt{\Omega_{m0}(1+z)^3+\Omega_{\Lambda0}+\Omega_{r_c0}}} \nonumber \\  &+&   \frac{{\Omega_{K0}(1+z)^2}}{(\sqrt{\Omega_{m0}(1+z)^3+\Omega_{\Lambda0}+\Omega_{r_c0}}-\sqrt\Omega_{r_c0})^2+\Omega_{K0}(1+z)^2}-1
\end{eqnarray}
and the transition redshift can be expressed as
\begin{equation}\label{zt2}
z_t=\frac{((4\Omega_{m0}\Omega_{r_c0}+2\Omega_{m0}\Omega_{\Lambda0}+2\sqrt{\Omega_{m0}^2\Omega_{r_c0}(4\Omega_{r_c0}+3\Omega_{\Lambda0})})\Omega_{m0})^{1/3}}{\Omega_{m0}}-1.
\end{equation}
One can check that this expression is exactly similar to the one obtained in a flat $\Lambda$DGP model. Now, we use the numeric $\chi^2$ method to obtain the best-fitted values of our model parameters. To this aim we have utilized the observational data from Type Ia Supernova (SNeIa), Baryon Acoustic Oscillations (BAO) and Cosmic Microwave Background radiation (CMB). To constrain our model parameters with attention to SNeIa, we use 557 data points belong to the Union sample \cite{Amanullah10}. The related $\chi^2$ value is defined as
\begin{equation}\label{chi2}
    \chi^2_{SNe}=\sum_{i=1}^{557}\frac{[\mu_i^{the}(z_i) - \mu_i^{obs}(z_i)]^2}{\sigma_i^2}.
\end{equation}
Here, $\mu_i^{the}$ and $\mu_i^{obs}$ are the theoretical and observational values of distance modulus parameter, respectively and $\sigma_i$ shows the observational estimated error. The distance modulus is the difference between the absolute and apparent magnitude of a distant object and is given by $\mu(z) = 5\log_{10}D_L(z) - \mu_0$ where $\mu_0 = 5 log_{10}h + 42.38$, $h = (H_0/100)$km/s/Mpc and
\begin{equation}\label{dl}
D_{L}(z)\equiv(1+z)\int_0^z{\frac{dz'}{E(z')}},
\end{equation}
is called the luminosity distance. Furthermore, we can constrain our model free parameters using the definition of BAO distance
\begin{equation}\label{bao}
    D_V(z_{BAO})=[\frac{z_{BAO}}{H(z_{BAO})}(\int_0^{z_{BAO}}\frac{dz}{H(z)})^2]^{1/3}.
\end{equation}
To this aim we use the joint analysis of the 2dF Galaxy Redshift Survey at $z = 0.20$ and SDSS data at $z = 0.35$ \cite{Reid10},\cite{Percival10} as the BAO distance ratio which is model independent
\begin{equation}\label{drbao}
   \frac{D_V(z=0.35)}{D_V(z=0.20)}=1.736\pm0.065.
\end{equation}
The related $\chi^2$ value can be obtained using
\begin{equation}\label{chibao}
    \chi^2_{BAO}=\frac{[(D_V(z=0.35)/D_V(z=0.20))-1.736]^2}{0.065^2}\cdot
\end{equation}
The CMB shift parameter, $R$, \cite{Wang06},\cite{Bond97} which defines as
\begin{equation}\label{r}
    R\equiv\Omega_{m0}^{1/2}\int_0^{z_{CMB}=1091.3}\frac{dz'}{E(z')},
\end{equation}
contains the major observational information from CMB. Thus to take into account the contribution of CMB in our analysis we use the $\chi^2$ below
\begin{equation}\label{chicmb}
    \chi^2_{CMB}=\frac{[R-R_{obs}]^2}{\sigma_R^2},
\end{equation}
where $R_{obs} = 1.725\pm0.018$ \cite{Komatsu11}. Now, minimizing $\chi^2_{SNe}+\chi^2_{BAO}+\chi^2_{CMB}$, we obtain the best-fit values of our model parameters. Also, in this procedure we have considered $z_t$, as a free parameter. The results have been shown in Table \ref{table:1}. In comparison with the corresponding transition redshift of $\Lambda$CDM model, $z_t=0.632$ \cite{Ade13}, in non-flat $\Lambda$DGP model the Universe starts its accelerated expansion phase, earlier. Note that the value of $\Omega_{r_c0}$, has been obtained indirectly, using the best-fitted values of other model parameters, together with the relation (\ref{zt2}).

\begin{table}[ht]
\caption{best-fitted values of non-flat $\Lambda$DGP model parameters} 
\centering 
\begin{tabular}{c|c} 
\hline\hline 
model parameters  &  best-fitted values \\ [4ex] 
\hline 
$\Omega_{m0}$ &  $0.291^{+0.001}_{-0.003}$ \\
\hline 
$z_t$ & $0.638^{+0.001}_{-0.003}$\\
\hline 
$\Omega_{K0}$ & $0.011^{+0.001}_{-0.002}$\\
\hline 
$\Omega_{\Lambda0}$ & $0.671^{+0.004}_{-0.002}$\\
\hline\hline
$\Omega_{r_c0}$ & $0.0001^{+0.0001}_{-0.0002}$\\
\hline
\end{tabular}
\label{table:1} 
\end{table}\

\begin{figure}[t]
\centering
\includegraphics[width=0.45\textwidth]{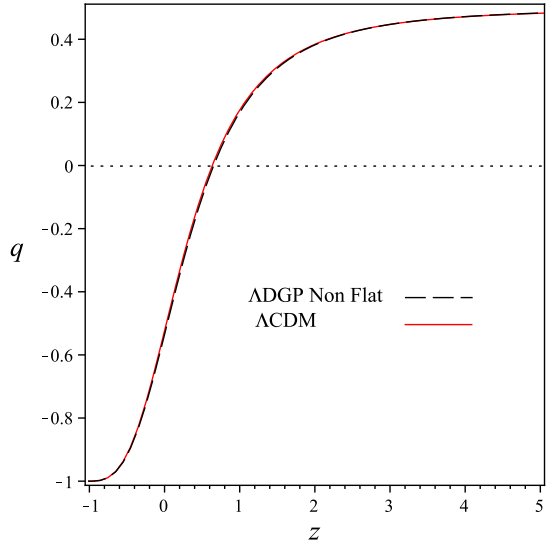}
\caption{Deceleration parameter as a function of redshift for $\Lambda$CDM model and non-flat $\Lambda$DGP model for the best-fitted values of our model parameters. The curve and the value of $z_t$, of two models are very similar to each other, but our model switches deceleration to acceleration a little earlier than $\Lambda$CDM model.}
\label{fig:q}
\end{figure}

\begin{figure}[t]
\centering
\includegraphics[width=0.45\textwidth]{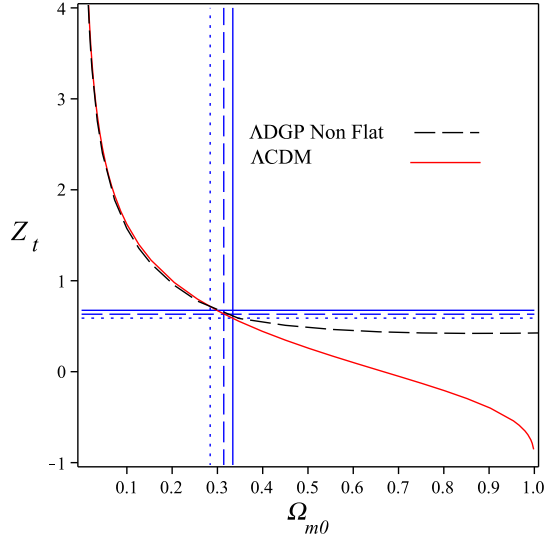}
\caption{Transition redshift as a function of $\Omega_{m0}$, for $\Lambda$CDM model and non-flat $\Lambda$DGP model for the best-fitted values of our model parameters. The horizontal and vertical blue lines are related to observations. The Planck satellite results in 68\% confidence limit on the present matter density parameter is $\Omega_{m0} = 0.314 \pm 0.020$ \cite{Ade13}. The best-fit value of the transition redshift and corresponding 68\% confidence limit, with attention to some observational data, has been obtained in \cite{Lu11}, using a parameterized deceleration parameter as $z_t= 0.69^{+0.20}_{-0.13}$. Both models are in good agreement with observations.}
\label{fig:zt}
\end{figure}

In Fig. \ref{fig:q}, the deceleration parameter as a function of redshift for best-fitted values, has been drawn with a dash curve in black. In Fig. \ref{fig:zt}, we have indicated transition redshift, $z_t$, as a function of $\Omega_{m0}$. Also, the observational data for $z_t$ and $\Omega_{m0}$, extracted from Planck satellite, have been shown by horizontal and vertical lines, respectively. It seems that, this model fits observations, well.

\section{REDSHIFT DRIFT}

\begin{figure}[t]
\centering
\includegraphics[width=0.45\textwidth]{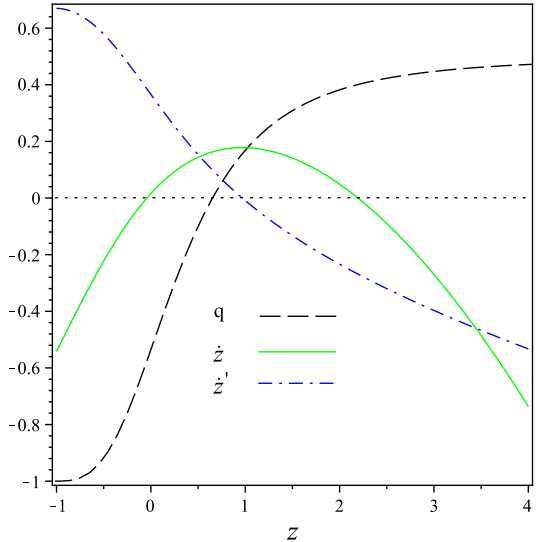}
\caption{$q$, $\dot z$ and $\dot z'$, as a function of redshift in non-flat $\Lambda$DGP model for the best-fitted values of our model parameters.}
\label{fig:qzdot}
\end{figure}

The cosmic redshift parameter is related to the scale factor, $a(t)$ and consequently to the cosmic time, $t$, indirectly. Then, we conclude that the redshift changes with time, as well. The variation of redshift with respect to time, called redshift drift, can be used to measure the variation of Hubble parameter, $H(z)$ and also the acceleration of the Universe, directly. According to \cite{Linder97}, if we introduce $t_0$ and $t_e$, respectively as the time in which a signal is detected in the frame of observer and the time in which the signal is emitted from the source, then
\begin{equation}\label{zdot}
    \frac{dz}{dt_0}=\frac{d}{dt_0}\left(\frac{a(t_0)}{a(t_e)}-1\right) =\frac{\dot a(t_0)-\dot a(t_e)}{a(t_e)}
\end{equation}
and one can reach to McVittie equation
\begin{equation}\label{Mc}
    \dot z=H_0(1+z)-H(z).
\end{equation}
The appearance of a difference in the value of $\dot a$, in (\ref{zdot}), is the reason that we can consider the redshift drift as a criterion of acceleration of the Universe. Also, using the best-fitted values in Table \ref{table:1} and with attention to McVittie equation, we can plot the curve $\dot z(z)$ and deal with the expansion history of the Universe. (See Fig. \ref{fig:qzdot}). We should note here that recently, the forecasting analysis, using the redshift drift has attracted a great deal of attention. It has been shown in a number of works that the mock redshift drift data can significantly improve the constraints on model parameters \cite{Corasaniti07},\cite{Geng14},\cite{Geng15}.

Using some calculations one can reach to relation below
\begin{equation}\label{qzdot}
    q=\frac{1-{\dot z}'}{1-\frac{\dot z}{1+z}}-1,
\end{equation}
where we have shown derivative with respect to $z$, by the prime. Also, from (\ref{Mc}), we can obtain
\begin{equation}\label{slope}
    \dot z'=H_0(1-E'(z)).
\end{equation}
With attention to (\ref{slope}), in the diagram $\dot z(z)$, the slope of the curve is negative, if $E'(z)>1$ and is positive, if $E'(z)<1$. Also, the extremum point is related to $E'(z)=1$. Analyzing equations (\ref{Mc}), (\ref{qzdot}) and (\ref{slope}) and with attention to Fig. \ref{fig:qzdot}, one can find that during decelerated expansion, $\dot z$ changes from negative to positive values. Then, starting the accelerating phase it turns and approach zero at the present and continues to negative values in the future. Furthermore, as it has been shown in Fig. \ref{fig:qzdot}, that $\dot z'$, changes from negative to positive, as well in decelerating phase. But during accelerated expansion, it extends to the larger positive values.

In a Universe filled only with matter component, the expansion is slowing down because of the effect of gravity. But today, we know from observations that our Universe at an epoch in its expansion history has started to accelerate where ensures the existence of a dark energy component with negative pressure. In terms of dimensionless density parameters of dark matter, $\Omega_m$ and dark energy, $\Omega_d$, this epoch is related to the redshift $z_{eq}$, in which $\Omega_m=\Omega_d$. This redshift differs from $z_t$, in which the Universe starts its acceleration. The comparison between these two values is another interesting object in cosmology. The corresponding value of $z_{eq}$, which can be obtained from \cite{Ade13}, shows $z_{eq}=0.298$, for $\Lambda$CDM model. In non-flat $\Lambda$DGP model and for simplicity we use new variables $\Omega_i'=E^2\Omega_i$. Then, $\Omega_m'=\Omega_{m0}(1+z)^3$ and $\Omega_d'=\Omega_{\Lambda0}$. The condition $\Omega_m'(z_{eq})=\Omega_d'(z_{eq})$, leads to an expression for $z_{eq}$, in terms of our model parameters as
\begin{equation}\label{zeq4}
z_{eq}=(\frac{\Omega_{\Lambda0}}{\Omega_{m0}})^{\frac{1}{3}}-1.
\end{equation}
Using Table \ref{table:1}, we obtain $z_{eq}=0.321$, in our model which is larger than the corresponding value in $\Lambda$CDM model. But, similar to the $\Lambda$CDM model, $z_t>z_{eq}$. Thus, without need to domination of dark energy component the Universe can start its accelerating expansion phase.

\begin{figure}[t]
\centering
\includegraphics[width=0.45\textwidth]{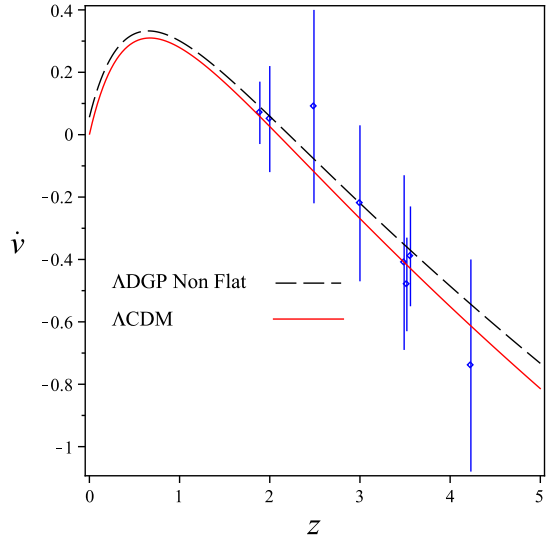}
\caption{Drift velocity as a function of redshift for $\Lambda$CDM model and non-flat $\Lambda$DGP model for the best-fitted values of our model parameters. Both models fit observations well.}
\label{fig:vdot}
\end{figure}

Besides, we know from observations that the velocity of a light source changes with respect to $t_0$, though this variation is very small. It is just in order of a few cm s$^{-1}$, if we consider a time interval about 30 years. There is a relation between this velocity drift and the redshift drift parameter as
\begin{equation}\label{vdot}
\dot v=\frac{ dv(z)}{dt_0}=\frac{ c}{(1+z)}\frac{dz}{dt_0},
\end{equation}
which is in order of several cm s$^{-1}$ yr$^{-1}$. Fig. \ref{fig:vdot}, shows the behavior of $\dot v(z)$, of both $\Lambda$CDM and $\Lambda$DGP model in comparison with observational data from CODEX, including eight points \cite{CODEX07}. Both the curves are in good agreement with observations.

\section{statefinder diagnostic}

To distinguish and classify different dark energy models, a few approach have been proposed. Among them, the statefinder diagnostic is of particular interest. This approach has been found in terms of two new geometrical variables which are related to the third derivative of scale factor with respect to time \cite{Sahni03}. In a non-flat Universe these two new variables define as

\begin{equation}\label{rs}
    r=\frac{\dot{\ddot a}}{aH^3}, \quad s=\frac{r-1+\Omega_K}{3(q-1/2+\Omega_K/2)}\cdot
\end{equation}

\begin{figure}[t]
\centering
\includegraphics[width=0.45\textwidth]{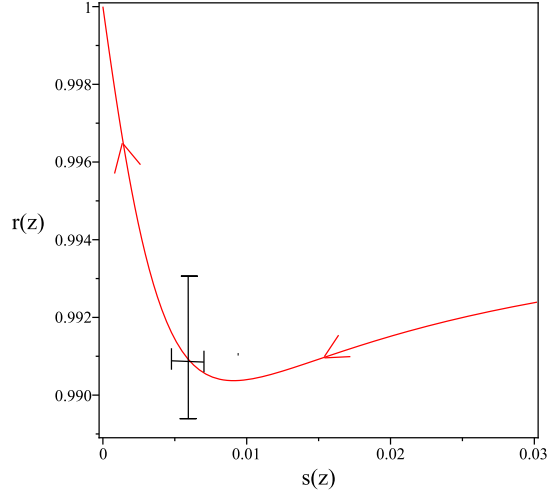}
\caption{The evolution of the statefinder parameter $r$ versus $s$, in non-flat $\Lambda$DGP model for the best-fitted values of our model parameters. There is a very small deviation from the point $(1,0)$, related to $\Lambda$CDM model. This confirms analogue and closeness of the two models.}
\label{fig:rs}
\end{figure}

\begin{figure}[h]
\centering
\includegraphics[width=0.45\textwidth]{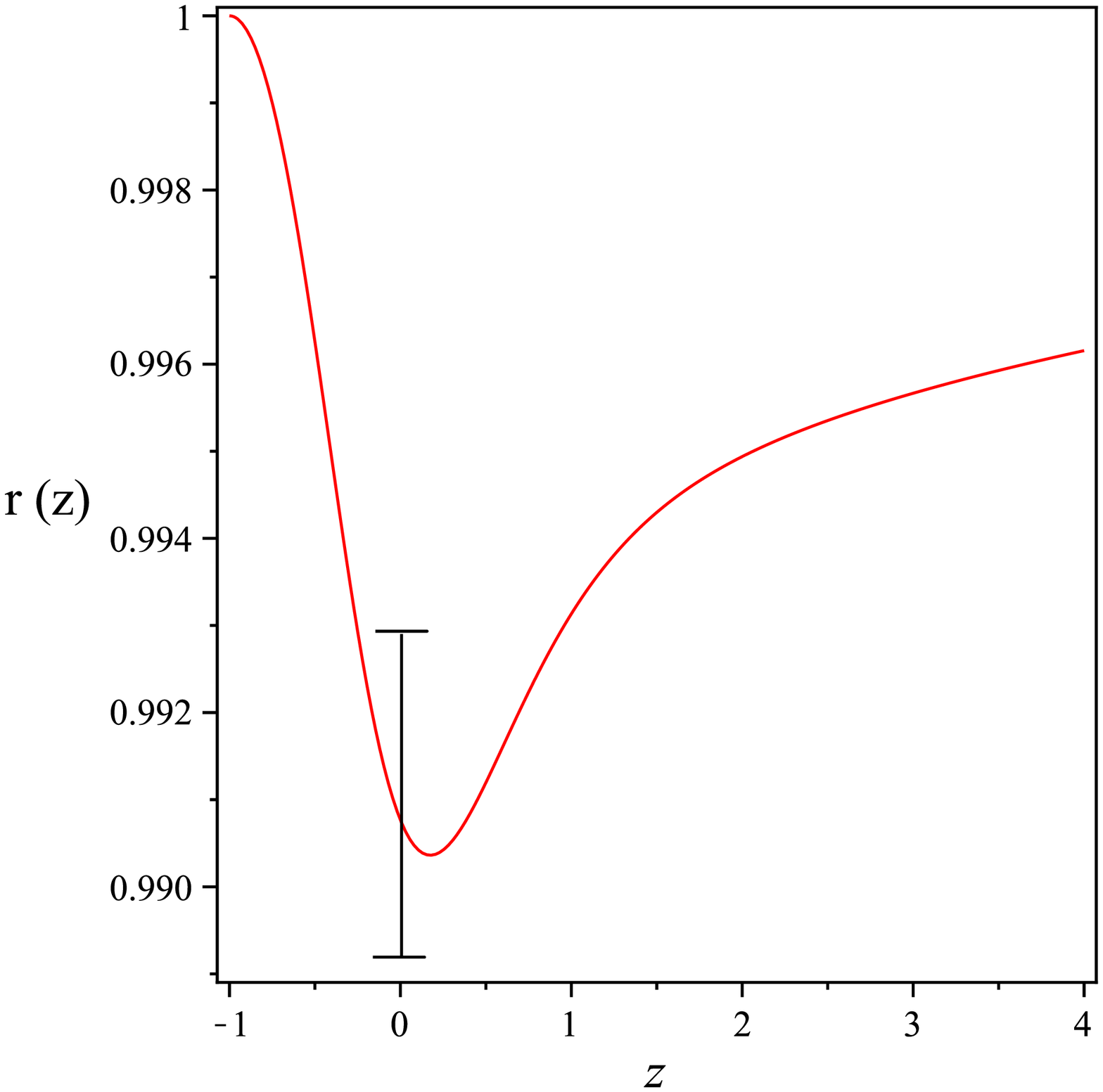}
\includegraphics[width=0.45\textwidth]{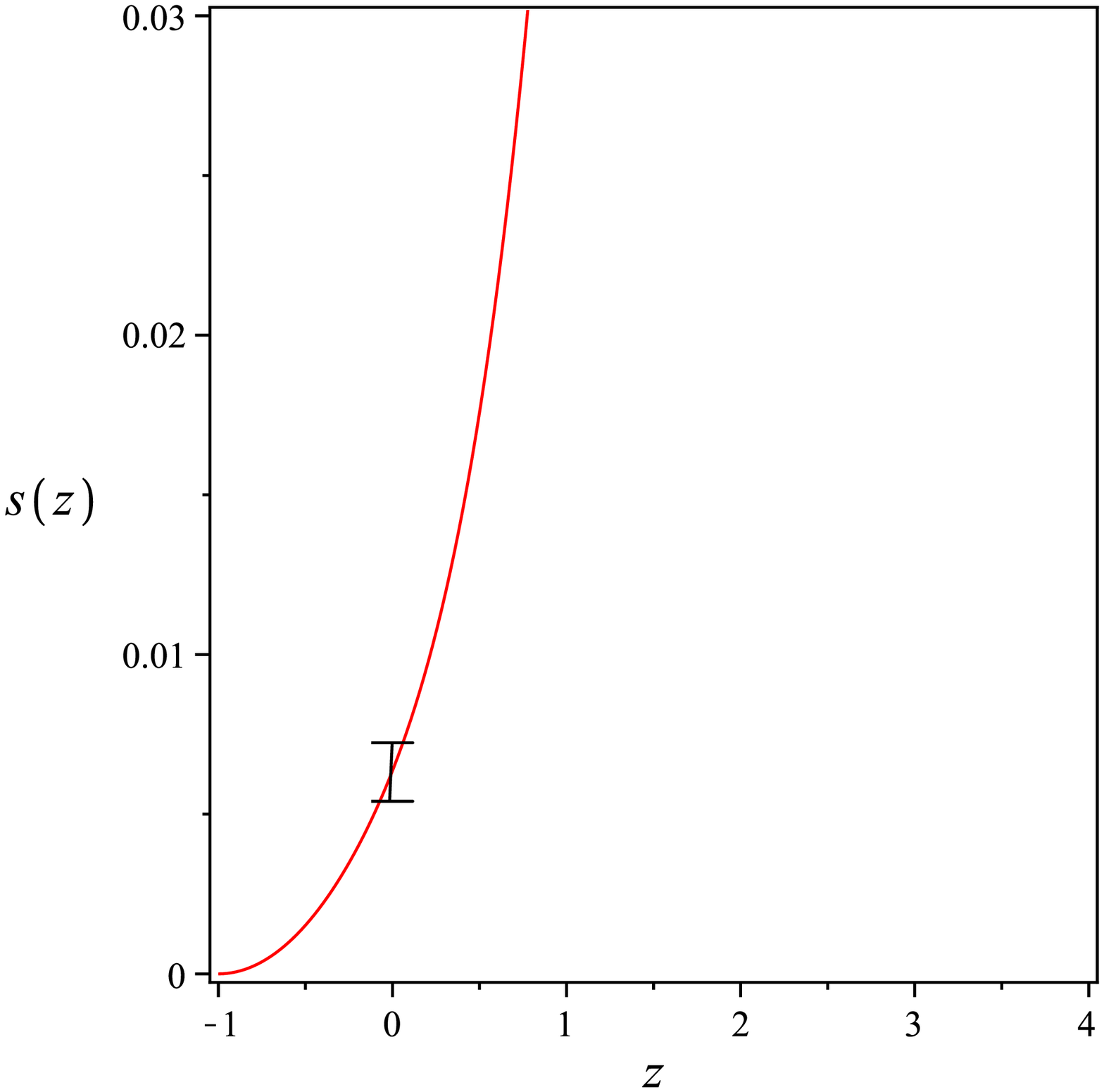}
\caption{The evolution of the statefinder parameters $r$ and $s$ versus redshift, in non-flat $\Lambda$DGP model for the best-fitted values of our model parameters.}
\label{fig:rands}
\end{figure}

Also, they can be rewritten in terms of the equation of state parameter, $w$, and its first derivative with respect to time \cite{Sahni03}, as
\begin{equation}\label{rsw}
    r=1-\Omega_K+\frac{9}{2}w_d(1+w_d)\Omega_d-\frac{3}{2}\frac{\dot w_d}{H}\Omega_d, \quad s=1+w_d-\frac{1}{3}\frac{\dot w_d}{w_dH}.
\end{equation}
So, for the $\Lambda$CDM model, with $w_d=-1$, we have $(r,s)=(1,0)$. The pair $(r,s)$, has been utilized frequently in the literature to discriminate a wide variety of dark energy models \cite{Alam03},\cite{Visser04},\cite{Zhang08},\cite{Zhang10},\cite{Setare11},\cite{Sami12},\cite{Cui14}. To this aim one can compare the corresponding trajectories in $r-s$ plane. Moreover, deviation from the fixed point $(1,0)$, related to the $\Lambda$CDM model can be studied using these curves. Fig. \ref{fig:rs}, illustrates the trajectories belong to $\Lambda$DGP model. The range of change of statefinder parameters, specially $r$, is very small, as it can be seen from Fig. \ref{fig:rands}, which means that our model has a tiny departure from $\Lambda$CDM model. Also, the curve $r(s)$, approaches the fixed point $(1,0)$ at late times.

\section{coincidence problem}

One of the most important problems of $\Lambda$CDM model is coincidence problem, namely, why the energy densities of dark matter and dark energy are of precisely the same order today? In another words if we introduce
\begin{equation}\label{coincidence}
    R=\frac{\rho_m}{\rho_d},
\end{equation}
then, the coincidence problem asks why $R$, is of order unity now?

Many scenarios have been proposed to solve or at least alleviate this problem. For instance, some dynamical dark energy models have been put forward to replace the cosmological constant \cite{Peebles03},\cite{Padmanabhan03}. Also, the coupling and interacting between dark sectors of the Universe has been used to this aim \cite{Chimento03},\cite{Olivares06},\cite{Amendola06},\cite{Campo06},\cite{Olivares08},\cite{Karwan08},\cite{Egan08},\cite{Kim08},\cite{Zhang09}. In these articles the authors have been investigated different approached to resolve the coincidence problem. Some of them show that $R$, is independent of initial conditions and study attractor solutions. Some others discuss $R$, does not change much during the whole history of the universe. Also, in many of them the authors introduce a mechanism in which $R$, tends to a constant value at late times or varies slower than the scale factor today.

\begin{figure}[t]
\centering
\includegraphics[width=0.45\textwidth]{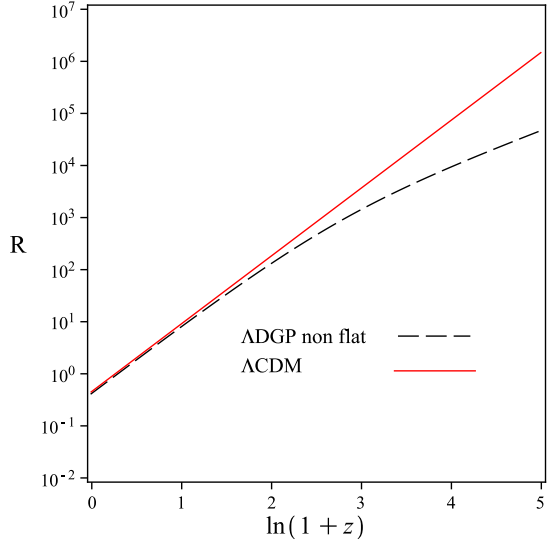}
\caption{The trajectory of $R$, versus $\ln(1+z)$, for $\Lambda$CDM model and non-flat $\Lambda$DGP model for the best-fitted values of our model parameters. The coincidence problem in our model has been alleviated.}
\label{fig:coin}
\end{figure}

\begin{figure}[t]
\centering
\includegraphics[width=0.45\textwidth]{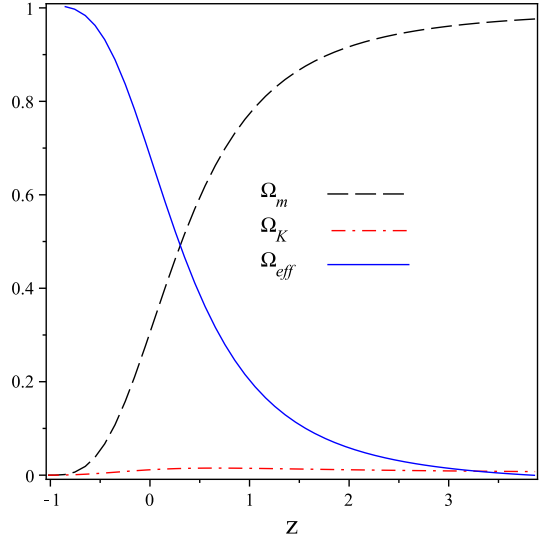}
\caption{The trajectory of $\Omega_m$, $\Omega_{eff}$ and $\Omega_K$, versus redshift, in non-flat $\Lambda$DGP model for the best-fitted values of our model parameters. It is obvious that $\Omega_m+\Omega_{eff}+\Omega_K=1$.}
\label{fig:omega}
\end{figure}

But in here, we does not replace $\Lambda$, with a dynamical dark energy term. Also, we does not consider any interaction between dark sectors. We are only trying to investigate the effect of the extra dimensions. We can show that in a $\Lambda$DGP model, though the coincidence problem does not solve in full, but it can be at least ameliorated. To this aim we can introduce an effective dark energy term in our model if we rewrite the Friedman equation in standard general relativistic from as $\Omega_m+\Omega_{eff}+\Omega_K=1$. Thus we obtain
\begin{equation}\label{Friedeff}
    \Omega_{eff}=\Omega_\Lambda-2\sqrt{\Omega_{r_c}}\sqrt{1-\Omega_K}
\end{equation}
and we can interpret the ratio (\ref{coincidence}), in our model as $R=\rho_m/\rho_{eff}$. Fig. \ref{fig:coin}, illustrates the behavior of $R$, in the whole history of the Universe until now for both $\Lambda$CDM and $\Lambda$DGP models. It is obvious that in our model the coincidence problem has been a little alleviated and this is only because of considering the effect of extra dimension. Also, Fig. \ref{fig:omega}, shows the behavior of $\Omega_m$, $\Omega_{eff}$ and $\Omega_K$ versus redshift in our model.

\section{conclusion}

In this paper we used a non-flat $\Lambda$DGP model and obtained best-fitted values of transition redshift, $z_t$, and other model parameters using SNe+BAO+CMB data. We found that transition from decelerating expansion to accelerating phase in our model happens earlier than in the $\Lambda$CDM model. The cosmic redshift drift studied exactly and the correlations between $q$, $\dot z$ and $\dot z'$ investigated in this model. We obtained $z_{eq}$, in our model and understood that like the $\Lambda$CDM model, before domination of dark energy component the Universe starts its accelerating expansion.

With attention to Figs. \ref{fig:zt} and \ref{fig:vdot}, we concluded that our model is in a good agreement with observational data released by Planck and CODEX, respectively. we see that our model is marginally consistent with transition redshift derived indirectly from observation, better than $\Lambda$CDM model. We exerted a statefinder diagnostic scenario in our model and found that this model is a tiny deviation from the $\Lambda$CDM model. Also, from Fig. \ref{fig:coin}, we found that our model improves the coincidence problem.

\label{lastpage}

\end{document}